\documentstyle[preprint,aps]{revtex}
\begin{document}
\preprint{KAIST-TH 19/96\hspace{.3cm} CBNU-TH 961108 \hspace{.3cm}
          hep-ph/xxxxxxx}

\title{\bf Proton Decay with  a Light Gravitino or Axino
}

\author{Kiwoon Choi${}^{\dagger}$ , Eung Jin Chun${}^*$,  
and Jae Sik Lee${}^{\dagger}$}

\address{Department of Physics, Korea Advanced Institute of Science
and Technology, Taejon 305-701, Korea${}^{\dagger}$}


\address{Department of Physics, Chungbuk National
University, Cheongju 360-763, Korea${}^*$}



\maketitle

\begin{abstract}

We consider the proton decay in supersymmetric models
with a  gravitino or axino lighter than the proton.
This consideration leads to a stringent limit on
the $R$ parity and $B$ violating Yukawa coupling 
of the superpotential operator $U^c_iD^c_jD^c_k$  
as $\lambda^{\prime\prime}_{112}\leq 
10^{-15}(m_{3/2}/{\rm eV})$ for a light gravitino,
and $\lambda^{\prime\prime}_{112}\leq 10^{-15}
(F_a/ 10^{10} \, {\rm GeV})$ for a light Dine-Fischler-Srednicki-
Zhitnitskii axino. For hadronic axino, the constraint is weakened 
by the factor of $10^{3}$.

\end{abstract}
\pacs{11.30.Fs, 12.60.Jv, 13.30.-a, 14.80.Mz}
\def\be{\begin{equation}}
\def\ee{\end{equation}}
\def\bea{\begin{eqnarray}}
\def\eea{\end{eqnarray}}

Proton stability  
strongly constrains  the baryon ($B$) and lepton  ($L$) number violating
couplings. Since all  known fermions lighter than the proton
carry a nonzero lepton number, the couplings (or the combinations
of couplings)  relevant
for the proton decay 
should conserve
$B-L$.
However if there is a lighter fermion which 
does {\it not}  carry any lepton number,
proton decay may be induced  
by  a $B$ violating but $L$ conserving
interaction alone \cite{chang}.
There are in fact very interesting class of 
models which predict
such a light fermion.
In supersymmetric models in which supersymmetry (SUSY)
breaking is mediated by  gauge interactions,
the squark and/or gaugino masses, i.e. the soft masses in the supersymmetric
standard model (SSM) sector,
are given by
$m_{\rm soft}\simeq ({\alpha\over \pi})^n \Lambda_S$
where $n$ is a model-dependent positive integer and
$\Lambda_S$ corresponds to the scale of spontaneous
SUSY breaking \cite{nelson}. 
In such models, in order for $m_{\rm soft}$ to be of order the weak scale,
$\Lambda_S$ is assumed to be $10\sim 1000$ TeV,
leading to the gravitino mass 
$m_{3/2}=\Lambda_S^2/M_P\leq 1$ keV  far below the proton mass.
If a global  $U(1)_{PQ}$ symmetry is introduced in 
a gauge-mediated SUSY breaking model
to solve the strong CP problem by the axion mechanism \cite{pq},
SUSY breaking in the axion sector is  mediated also by some 
gauge interactions.
The axino mass in such models  is given by
$m_{\tilde{a}}\simeq (\alpha/\pi)^m\Lambda_S^2/F_a$ where
$m$ is again a model-dependent  (but typically bigger than $n$)
positive integer and  $F_a$ denotes the scale
of spontaneous $U(1)_{PQ}$ breaking \cite{axino}.
Obviously then
the axino is lighter than the proton
for  a  phenomenologically allowed $F_a\geq 10^{10}$ GeV.
In other type of models in which SUSY breaking is transmitted
by supergravity interactions,  the gravitino mass is fixed to be
of order the weak scale, however there is still a room for an axino
lighter than the proton \cite{chun}.
As was pointed out in Ref. \cite{chun},
some supergravity-mediated models lead to
$m_{\tilde{a}}\simeq m_{3/2}(m_{3/2}/M_P)^{1/2}\simeq 1$ keV for which
the axino would be a good warm dark matter candidate \cite{turner}.
In this paper, we wish to examine the proton decay involving
a light gravitino or axino to derive a constraint
on the superpotential interaction  $\lambda^{\prime\prime}_{ijk}
U^c_iD^c_jD^c_k$
which violates  $R$ parity and $B$, while conserving $L$.

Let us first consider the proton decay involving a light gravitino,
more precisely the helicity $\pm 1/2$ Goldstino component.
Our starting point is the effective lagrangian below
the scale $\Lambda_S=\sqrt{m_{3/2}M_P}$   but above
the weak scale soft mass $m_{\rm soft}$:
\be
{\cal L}={\cal L}_{\rm SSM}+{\cal L}_{G},
\ee
where ${\cal L}_{\rm SSM}$ denotes the lagrangian density
of the SSM fields
and the  Goldstino lagrangian ${\cal L}_{G}$
is given by \cite{fayet}
\be
{\cal L}_{G}=\frac{i}{2}\bar{G}\gamma^{\mu}\partial_{\mu}G+
\frac{i}{4\sqrt{6}m_{3/2}M_p}
\left( \bar{\lambda}^a\gamma^{\rho}\sigma^{\mu\nu}
\partial_{\rho}GF_{\mu\nu}^a+ 
2\sqrt{2} \bar{\psi}_{I}(1-\gamma_5)
\gamma^{\mu}\gamma^{\nu}
\partial_{\mu}GD_{\nu}\phi^*_I+ {\rm h.c} \right)
\ee
where $G$ denotes  the four-component Majorana
Goldstino field.
Here
${\cal L}_{\rm SSM}$ includes the terms associated with
the $B$ violating superpotential interaction,
\be
W_{\rm SSM} \, \ni \, \lambda^{\prime\prime}_{ijk}U^c_iD^c_jD^c_k,
\ee
and $(\phi_I,\psi_I)$ and $(\lambda^a, F^a_{\mu\nu})$ stand for
the left-handed chiral matter and gauge multiplets in the SSM sector.
Note that the above form of Goldstino lagrangian is enough
for the study of the process involving a single
on-shell Goldstino obeying $i\gamma^{\mu}\partial_{\mu}G=m_{3/2}G$.

Integrating out all fields heavier than the scale of
the QCD chiral symmetry breaking, i.e. $\Lambda_{\chi}\simeq
1$ GeV, we are left with an effective lagrangian
of the light quarks, $q_{\alpha}$ ($\alpha = (u, d, s)$),
and gluons
together with the light Goldstino
(of course also 
the light leptons and the photon which are not relevant for
our discussion).
The operators responsible for the proton decay in this  
effective lagrangian at $\Lambda_{\chi}$
are induced by the exchange of the $SU(2)_L$ singlet squarks as 
\be
{\cal O}_{\rm eff}=\frac{2i\lambda^{\prime\prime}_{112}y_{\alpha\beta\gamma}}
{\sqrt{3}m_0^2m_{3/2}M_P}
(\bar{q}_{\alpha}(1-\gamma_5)q^c_{\beta})(\partial_{\mu}\bar{q}_{\gamma}
(1-\gamma_5)\partial^{\mu}G).
\ee
Here $m_0^2$ denotes the squark masses which are assumed to be
(approximately) universal, 
\be
y_{dsu}=y_{uds}=
y_{usd}=1,
\ee
and all other components of $y_{\alpha\beta\gamma}$ do vanish. 
Note that  the above  operator has $B=S=-1$,
and thus the relevant proton decay mode is  $p\rightarrow G+K^+$.
For a generic non-universal squark mass matrix,
$S=0$ operator can be induced also to give rise to 
$p\rightarrow G+\pi^+$, however it 
is suppressed by a small squark mixing. 
To arrive at the above interaction operator, we 
have used the equation of motion of the on-shell Goldstino field
and ignored
the piece suppressed by the small $m_{3/2}$. Also ignored
are the renormalization effects between the weak scale and
$\Lambda_{\chi}$.

The hadronic matrix elements  of the  above 
$B=S=-1$ operator would be
described by an effective chiral lagrangian including the Goldstino
field. 
Let us  consider a chiral operator ${\cal O}_{\chi}$ 
which would  induce $p\rightarrow G+K^+$ as
a low energy realization of
the light quark operator ${\cal O}_{\rm eff}$ below $\Lambda_{\chi}$.
Obviously it  can
be written as ${\cal O}_{\chi}=Z^{\mu}(1-\gamma_5)
\partial^{\mu}G$ where $Z^{\mu}$ is a fermionic $B=S=-1$  operator including
$\bar{P}$ and $K^{+}$.
If $Z^{\mu}$ does not include any spacetime derivative,
${\cal O}_{\chi}$ is suppressed by the small factor
$m_{3/2}/m_p$ 
(for on-shell Goldstino) where $m_p$ denotes the proton mass.
For $Z^{\mu}$ containing a single spacetime derivative,
we have 
\be
{\cal O}_{\chi}=
\frac{2\xi \lambda^{\prime\prime}_{112}}
{\sqrt{3} m_0^2m_{3/2}M_P}(\bar{P}(1-\gamma_5) 
\partial^{\mu}G)\partial_{\mu}K^+,
\ee
where again the equations of motion are used together with
$m_{3/2}\ll m_p$.
To estimate the size of the hadronic coefficient $\xi$,
we use the naive dimensional analysis (NDA) rule of Ref. \cite{manohar},
yielding
\be
|\xi|\simeq 4\pi f_{\pi}^2,
\ee
where $f_{\pi}=93$ MeV is the pion decay constant.
In fact, the NDA rule gives
$\Lambda_{\chi}=4\pi f_{\pi}$ and then
the typical energy in the proton decay, i.e.
$m_p$, is
comparable to  $\Lambda_{\chi}$.
This means that, within the NDA rule,
chiral operators with more spacetime
derivatives are equally important as the operator
of  Eq. (6).
However for 
an order of magnitude estimate
of the hadronic matrix element,  the consideration of
$Z^{\mu}$  with a single derivative would be 
enough.
Then applying the experimental limit on 
$p\rightarrow K^++\nu$ for $p\rightarrow K^++G$ 
induced by the interaction of Eq. (6),
we find the following constraint on the $R$ parity and $B$ violating
coupling:
\be
\lambda^{\prime\prime}_{112}\leq 5\times 10^{-16}
\left(\frac{m_0}{300 \, {\rm GeV}}\right)^2
\left(\frac{4\pi f_{\pi}^2}{|\xi|}\right)
\left(\frac{m_{3/2}}{1 \, {\rm eV}}\right), 
\ee
which is one of the main results of this paper.

Let us now consider the proton decay involving a light axino.
Similarly to the case of a light gravitino,
we start from the effective lagrangian at scales below
the scale $F_a$ of $U(1)_{PQ}$ breaking
but above $m_{\rm soft}$:
\be
{\cal L}={\cal L}_{\rm SSM}+{\cal L}_A,
\ee
where the axino lagrangian ${\cal L}_A$ can be read off from
\be
\int d^2\theta d^2\bar{\theta}
\, \frac{c_{_I}}{F_a}(A+{A}^{\dagger})\Phi_I^{\dagger}\Phi_I+
\left\{ \int d^2 \theta \, {c_a\over 16\pi^2 F_a}
AW^aW^a+{\rm h.c} \right\},
\ee
where $A=(s+ia)+\sqrt{2}\theta\tilde{a}+F_A\theta^2$
is  the axion superfield containing the axion $a$,
the saxion $s$ and the axino $\tilde{a}$,
while $\Phi_I$ and $W^a$
are the chiral superfields for the SSM matter and gauge
multiplets $(\phi_I,\psi_I)$ and $(\lambda^a, F^a_{\mu\nu})$, respectively. 
Here $c_{_I}$ and $c_a$ are dimensionless real coefficients.
For $F_a$ defined as the scale
of spontaneous $U(1)_{PQ}$ breaking, the 
coefficients $c_a$ of the axion coupling to the gauge multiplets
are of order unity in general.
However as we will discuss later,
the size of
the coefficients $c_{_I}$
of the axion coupling to the matter multiplets is somewhat
model-dependent.
Note that the above lagrangian
corresponds to the supersymmetric generalization 
of the conventional axion effective lagrangian \cite{kim1}:
\be
{\cal L}_a=\frac{2c_{_I}}{F_a}
\partial_{\mu}a\bar{\psi}_{I}\gamma^{\mu}\gamma_5\psi_{I}
+\frac{c_a}{32\pi^2 F_a}aF^{a\mu\nu}\tilde{F}^a_{\mu\nu}.
\ee
Obviously it is manifestly  
invariant under the nonlinear
$U(1)_{PQ}$ transformation, $A\rightarrow
A+ic$ ($c=$ real constant), up to the PQ anomaly.
At any rate, the relevant axino lagrangian is given by 
\bea
{\cal L}_A=&&
{1\over 2}i\bar{\tilde{a}}\gamma^{\mu}\partial_{\mu}\tilde{a}
-\frac{c_{_I}}{2F_a}(i
\partial_{\mu}\bar{\psi}_I\gamma^{\mu}(1+\gamma_5)\tilde{a}\phi^*_I
+{\rm h.c}) \nonumber \\
&& +\frac{c_a}{32\sqrt{2}\pi^2 F_a}
(\bar{\lambda}^a\gamma^{\mu}\gamma^{\nu}(1-\gamma_5)\tilde{a}F^a_{\mu\nu}+
{\rm h.c}),
\eea
where $\tilde{a}$ denotes the four-component Majorana
axino field.
Again the exchange of the $SU(2)_L$ singlet squarks
leads to  the following  $B=S=-1$ interaction  
in the effective lagrangian at $\Lambda_{\chi}$:
\be
{\cal O}_{\rm eff}=\frac{i\lambda^{\prime\prime}_{112}
y_{\alpha\beta\gamma}c_{\gamma}}{m_0^2F_a}
(\bar{q}_{\alpha}(1-\gamma_5)q^c_{\beta})\partial_{\mu}
\bar{q}_{\gamma}\gamma^{\mu}
(1+\gamma_5)\tilde{a},
\ee
where $c_{\gamma}$ ($\gamma=u,d,s$) denotes the axino coupling to 
the supermultiplet containing the $SU(2)_L$ singlet
right-handed light quark $q_{\gamma R}$
in Eq. (12) and
the squark degeneracy is assumed also.

Similarly to the gravitino case, in order to estimate
the  proton decay rate from the above effective interaction,
we consider a
chiral operator of the form ${\cal O}_{\chi}=X^{\mu}\gamma_{\mu}(1+\gamma_5)
\tilde{a}$ where $X^{\mu}$ is a $B=S=-1$
fermionic current made of $\bar{P}$ and $K^+$ which are on mass-shell.
For $X^{\mu}\propto K^+\bar{P}\gamma^{\mu}$,
the chiral operator ${\cal O}_{\chi}$
with the smallest number of spacetime derivatives is given by
\be
\frac{\xi_{\gamma} c_{\gamma}\lambda^{\prime\prime}_{112}}{m_0^2F_a}
(\bar{P}(1+\gamma_5)\tilde{a}) K^+,
\ee 
where the hadronic coefficients $\xi_{\gamma}$ 
are again determined by the NDA rule as
\be
|\xi_{\gamma}|\simeq 16\pi^2 f_{\pi}^3.
\ee
This then leads to the experimental bound on the
$R$ parity and $B$ violating coupling as
\be
\lambda^{\prime\prime}_{112}\leq 7\times10^{-16}
\left(\frac{m_0}{300 \, {\rm GeV}}\right)^2
\left(\frac{16\pi^2 f_{\pi}^3}{c_{\gamma}|\xi_{\gamma}|}\right)
\left(\frac{F_a}{10^{10} \, {\rm GeV}}\right),
\ee
which is another result of this paper.

The above constraint from the proton decay involving a light
axino depends upon
the dimensionless
coefficients $c_{\gamma}$ describing the axino coupling
to the supermultiplets of the $SU(2)_L$ singlet quarks [see Eq. (12)],
as well as the axion scale $F_a$.
In fact, the size of $c_{\gamma}$ has a certain model-dependence.
If the quark superfields carry a nonzero
$U(1)_{PQ}$ charge, which would be the case 
for the supersymmetric extension of the Dine-Fischler-Srednicki-Zhitnitskii
(DFSZ)
axion model \cite{dfsz}, 
the coefficients $c_{\gamma}$ would be of order unity in general.
However in hadronic axion models \cite{kim}
in which all SSM fields have a vanishing
$U(1)_{PQ}$ charge, the coefficients $c_{\gamma}$ are zero at tree level.
However the axino-quark couplings are
radiatively generated
through the axino coupling to the gluon multiplet,
yielding $c_{\gamma}\simeq
(\alpha_c/\pi)^2\ln(F_a/m_{\rm soft})\simeq 10^{-3}\sim 10^{-2}$
\cite{kim1}.
Thus the constraint  for hadronic axion models  becomes  weaker
than that for DFSZ models by the factor
of $10^2\sim 10^3$.

To conclude, we have considered the proton decay involving a 
gravitino or axino lighter than the proton. Generic models in which
supersymmetry breaking is mediated by gauge interactions contain
such a light gravitino. Then the $R$ parity and $B$ violating
coupling $\lambda^{\prime\prime}_{112}$ is strongly  constrained 
by the proton stability [see Eq. (8)]
to be less than about $10^{-15}(m_{3/2}/{\rm eV})$.
About  the possibility of a light axino,
gauge-mediated supersymmetry breaking models  endowed
with a global $U(1)_{PQ}$ symmetry generically predict an axino lighter
than the proton.
Also some supergravity-mediated models 
can give rise to a light axino, while the gravitino mass in such
models is fixed to be the weak scale.
We find  that  $\lambda^{\prime\prime}_{112}$ in models with
a light axino is constrained [see Eq. (16)] to be less than about
$10^{-15}(F_a/10^{10} \, {\rm GeV})$ and $10^{-12}(F_a/10^{10} \, {\rm GeV})$
for  Dine-Fischler-Srednicki-Zhitnitskii axion
models and hadronic axion models respectively.

\acknowledgements
This work is supported in part  
by KOSEF Grant 951-0207-002-2 (KC, JSL), KOSEF through CTP of Seoul
National University (KC), Programs of Ministry of Education
BSRI-96-2434 (KC), and
Non Directed Research Fund of KRF (EJC).  
EJC is a Brain-Pool fellow.

\end{document}